\documentclass[journal]{IEEEtran}
\usepackage{mathrsfs}
\usepackage{graphicx}
\usepackage{amssymb} 
\usepackage{amsmath} 
\usepackage{amsthm}
\usepackage{enumerate}
\usepackage{amsfonts}
\usepackage{bm}
\usepackage[numbers,sort&compress]{natbib}
\usepackage{color}

\newtheorem{theorem}{Theorem}
\newtheorem{remark}{Remark}

\newtheorem{example}{Example}
\newtheorem{lemma}{Lemma}
\newtheorem{definition}{Definition}

\usepackage{xcolor}  
\definecolor{emph_color}{rgb}{0,0,1}
\ifCLASSINFOpdf

\else
 \fi
\begin{document}
%
\title{A universal framework of GKYP lemma for singular fractional order systems}

\author{Yuman~Li,~
        Yiheng~Wei,~
        Yuquan~Chen,
		and~Yong~Wang

\thanks{Y. Li, Y. Wei, Y. Chen and Y. Wang are with the Department of Automation, University of Science and Technology of China, Hefei 230027, China. ~\tt\small{E-mail: lym2014@mail.ustc.edu.cn; neudawei@ustc.edu.cn; cyq@mail.ustc.edu.cn; yongwang@ustc.edu.cn}}

%
%
}


\maketitle

\begin{abstract}
The well-known GKYP is widely used in system analysis, but for singular systems, especially singular fractional order systems, there is no corresponding theory, {for which many control problems for this type of system can not be optimized in the limited frequency ranges.} In this {paper}, a universal framework of finite frequency band GKYP lemma for singular fractional order systems is established. Then the bounded real lemma in the sense of \({L_\infty }\) is derived for different frequency ranges. Furthermore, the corresponding controller is designed  to improve the \({L_\infty }\) performance index of singular fractional order systems. Three illustrative examples are given to demonstrate the correctness and effectiveness of the theoretical results.
\end{abstract}

\IEEEpeerreviewmaketitle

\section{Introduction}\label{Section 1}
{The singular fractional order system (SFOS) is a research hotspot in recent years. It plays an important role in many applications, such as optimization problems, economics,  constrained mechanics, biology, aircraft, robot dynamics, electric networks and large systems. Many scholars have paid attention to the analyses and control of SFOSs in many aspects, such as stability \cite{ndoye2010MED,ndoye2013Automatica,liu2017multiple}, admissibility \cite{marir2017JFI}, iterative learning control \cite{lazarevic2016JVC} and feedback control \cite{wei2017ISAtransactions,liu2017adaptive}. The $L_\infty$ norm of the transfer function matrix is an important performance index for SFOSs, which plays a significant role in energy calculation, controller design and filter analysis \cite{yin2015robust,shen2017stability,shen2014gain}. Generally speaking, there are many methods to solve $L_\infty$ norm, among which the most effective method is to establish the $L_\infty$ bounded real lemma by using KYP lemma and GKYP lemma \cite{liang2015bounded}.}

{The KYP lemma was first proposed in \cite{Rantzer1996KYP-lemma}, which effectively establishes the connection between time domain method and frequency domain method, and transforms the complex frequency domain inequality into a simple time domain linear matrix inequality. It enables comprehensive control of the system over the full frequency range. The KYP lemma was generalized to singular systems by Xu who gave a series of theories of positive real control for singular systems in \cite{xu:2006book}. Subsequently, there were many studies based on KYP lemma for singular systems \cite{camlibel2009SCL,reis2010IJC,feng:2015JFI} and people paid more and more attention to the research of singular systems \cite{xiong2017TSMCS,Liu2018TSMCS, wang2018TAC,wang2018Cyb,pang2018SMCS}. More recently, Bhawal et al. established a new KYP lemma for strong passive singular systems \cite{bhawal:2018IJC}.}

{Although the KYP lemma has been well generalized to singular systems, GKYP lemma has not. The GKYP lemma was first proposed by Iwasak in 2005 to solve control problems in different frequency ranges \cite{Iwasaki2005GKYPlemma}. Compared with KYP lemma, GKYP lemma is more practical, because in actual control problems, systems are usually required to meet different performance specifications in different frequency ranges. The proposal of GKYP lemma provided new ideas for many control problems, thus there are plenty of researches on GKYP lemma \cite{li2014heuristic,gao2011hinfty,hoang2008lyapunov,li2015robust}. Recently, the GKYP lemma has also been extended to fractional order systems  \cite{liang2015bounded,zhu2017FGKYP}. But for singular systems, the research in this area is remarkably lagging. In \cite{mei:2009CSSP}, Mei first designed the controller of singular perturbed systems based the GKYP lemma of non-singular systems by using the fast and slow subsystem decomposition method. Based on this idea, many researches on the finite frequency control of singularly perturbed systems have been developed \cite{huang2011IJCAS,zhai:2014JFI,xu2015JFI}. However, this kind of normalization method could only apply to a part of SFOSs which can be normalized to non-singular systems, and could not handle systems that do not satisfy the normalization condition. Up to now, there is no universal GKYP lemma for singular systems, regardless of fractional or integer-order systems. This leads to the fact that researches of singular integer and fractional order systems are limited to the whole frequency band, or only those systems which can be normalized into non-singular systems are able to be controlled in a limited frequency band.}

{Motivated by these findings, the GKYP lemma of SFOSs in the finite frequency ranges without normalization restriction is established in this paper, which can also be applied to singular integer order systems. Furthermore, the $L_\infty$ bounded real theorem and $L_\infty$ controller are obtained. The establishment of these theories will fundamentally solve the finite frequency control problem of singular systems, and provide a new perspective for other researches of SFOSs, like $H_\infty$ control and optimal tracking problem in finite frequency bands, the optimization of uncertain SFOSs in different frequency ranges and the decoupling of large-scale systems' singular representation.}

The rest of this {paper} is arranged as follows. {In Section \ref{Section 2}, some basic facts on this work are provided. Section \ref{Section 3} derives the universal framework of finite frequency band GKYP lemma for the SFOS. } In section \ref{Section 4}, $L_\infty$ bounded real lemma of SFOS is derived in different frequency ranges. Then an $L_\infty$ controller is designed in Section \ref{Section 5}. The numerical simulation is performed in Section \ref{Section 6}. Finally, conclusions in Section \ref{Section 7} close the {paper}.

\textbf{Notations.}
For matrix $X$, the symbols $ X^{\rm{T}}$ and $X^*$ represent the transpose and complex conjugate transpose, respectively. Expression $X > 0 \ (X < 0)$ indicates that $X$ is positive (negative) definite. The symbol ${\rm sym}(X)$ is an abbreviation for  $X+X^*$, and \({\delta _{\max }}(X)\) represents the maximum singular value of $X$. The symbols \({\mathbb{C}^{m \times n}}\) and \({\mathbb{R}^{m \times n}}\) stand for sets of $m \times n$ complex and real matrices, respectively. $\mathbb{H}_n$ denotes the set of $n \times n$ complex Hermitian matrices. The trace and rank of a matrix $X$ are represented by $\rm{tr}$$(X)$ and $\rm{rank}$$(X)$, respectively. ${\rm Re}(X)$ and ${\rm Im}(X)$ denote the real and imaginary parts of $X$. The operator $\otimes$ is the Kronecker's product. The convex hull and the interior of a set $S$ are denoted by $\rm{co}$$(S)$ and $\rm{int}$$(S)$, respectively. {The set $\mathbb{N}$ represents a set of matrices that satisfy $N=N^*\leq0$.}

\section{Preliminaries}\label{Section 2}
\subsection{Singular fractional order system  model}
Consider a singular fractional order system
\begin{equation}\label{Eq1}
\left\{ {\begin{array}{*{20}{c}}
{E{{\mathscr {D}}^\alpha }x\left( t \right) = Ax\left( t \right) + Bu\left( t \right)},\\
\ \ \ \ \ \ {y\left( t \right) = Cx\left( t \right) + Du\left( t \right)},
\end{array}} \right.
\end{equation}
{where $x\left( t \right)\in{\mathbb{R}^n}$ is the state variable of the system}, \(u\left( t \right) \in {\mathbb{R}^m}\) is the control input, \(y\left( t \right) \in {\mathbb{R}^p}\) is the {measured output}, $B, C$ and $D$ are constant matrices with appropriate dimensions, \({\mathscr {D}^\alpha }\) represents the Caputo fractional derivative, \(\alpha \) is the commensurate order of the SFOS and \({\rm{0 < }}\alpha {\rm{ < 2}}\),  \(E,A \in {\mathbb{R}^{n \times n}}\), and $\rm{rank}$$\left( E \right) = r < n$.

If the SFOS is relaxed at \(t = 0\), the transfer function matrix between \(u\left( t \right)\) and \(y\left( t \right)\) is
\begin{equation}\label{Eq2}
G(s) = C{\left( {{s^\alpha }E - A} \right)^{ - 1}}B + D.
\end{equation}

\subsection{S-procedure}
In deriving the process of GKYP lemma, S-procedure is a very important tool.
Given \(\Theta ,M \in {\mathbb{H}_q}\), if the regularity, $M>0$, is assumed, there exists the equivalence
\begin{equation*}
\begin{array}{l}
{\xi ^*}\Theta \xi  < 0,\forall \xi  \in \mathbb{C}^q \ \text{such that  }\xi  \ne 0,{\xi ^*}M\xi  \ge 0.\\
 \Leftrightarrow \exists \tau  \in \mathbb{R}\text{ such that  }\tau  \ge 0,\Theta  + \tau M < 0.
\end{array}
\end{equation*}
{To} generalize the above S-procedure, paper \cite{Iwasaki2005GKYPlemma} rewrites it with a different notation.
Define set \(\mathbb{S}_1\) specified by $\mathbb{M}$ as follows
\begin{equation}\label{Eq5}
\mathbb{S}\left( \mathbb{M} \right) = \left\{ {S \in {\mathbb{H}_q}:S > 0,{\rm tr}\left( \mathbb{M}S \right) \ge 0} \right\},
\end{equation}
\begin{equation}\label{Eq6}
{\mathbb{S}_1}\left( \mathbb{M} \right) = \left\{ {S \in \mathbb{S}\left( \mathbb{M}\right):\rm{rank} \left(\emph{S} \right) = 1} \right\},
\end{equation}
where
\begin{equation*}
\mathbb{M} = \left\{ {\tau M:\tau  \in \mathbb{R},\tau  \ge 0,M \in {\mathbb{H}_q}} \right\}.
\end{equation*}
Then, the S-procedure can be stated as
\begin{equation}\label{Eq4}
\rm{tr}(\Theta {\mathbb{S}_1}) < 0 \Leftrightarrow \left( {\Theta  + \mathbb{M}} \right) \cap {\mathop{\rm int}} (\mathbb{N}) \ne {\O}.
\end{equation}
Clearly, the S-procedure is completely specified by the set $\mathbb{M}$. {If in the equation (5), ``\( \Rightarrow \)"  and ``\( \Leftarrow \)"  are simultaneously established, the S-procedure is considered to be lossless. The lossless  condition of S-procedure has been demonstrated in \cite{Iwasaki2005GKYPlemma} as following.}

\begin{definition}\label{Definition 1}
\(\mathbb{M} \subset {\mathbb{H}_q}\) is said to be\\
i) admissible if it is a nonempty closed convex cone and \({\mathop{\rm int}} \left( \mathbb{N} \right) \cap \mathbb{M} = {\O} \);\\
ii) rank-one separable if \( \mathbb{S} = \rm{co}(\mathbb{S}_1)\).
 \end{definition}
 \begin{lemma}\label{Lemma 1}  \cite{Iwasaki2005GKYPlemma} Let \(\mathbb{M} \subset {\mathbb{H}_q}\) be defined by (3) and (4). Then for any matrix \(\Theta  \in {\mathbb{H}_q}\), if and only if the set $\mathbb{M}$ is admissible and rank-one separable, the strict S-procedure is lossless.
 \end{lemma}
\begin{remark}\label{Remark 1}
Lemma 1 shows that when $\mathbb{M}$ is chosen to be admissible and rank-one separable, no matter which \(\Theta \) is selected, the S-procedure will be lossless. This is important in the following subsequent proof process.
\end{remark}
\begin{lemma}\label{Lemma 2}\cite{Iwasaki2005GKYPlemma}
Let \(\mathbb{M} \subset {\mathbb{H}_q}\) be a rank-one separable set. Then
the set \({F^*}\mathbb{M}F + \mathbb{P}\) is rank-one separable for an arbitrary
\(F \in {\mathbb{C}^{q \times p}}\) and subset \(\mathbb{P} \subset {\mathbb{H}_p}\) of positive semi-definite matrices containing the origin.
\end{lemma}
\subsection{Frequency range}

\begin{definition}\label{Definition 2}
A curve on the complex plane is a collection of infinitely many points \(\lambda \left( t \right) \in {\mathbb{C}^{q \times p}}\) continuously parameterized by $t$ for \({t_0} \le t \le {t_f}\) where \({t_0},{t_f} \in \mathbb{R} \cup \left\{ { \pm \infty } \right\}\) and \({t_0} < {t_f}\). A set of complex numbers \(\Lambda  \subseteq \mathbb{C}\)  is said to represent a curve (or curves) if it is a union of a finite number of curve(s). With \(\Phi ,\Psi  \in {H_2}\) being given matrices, \(\Lambda \) is defined as
\begin{equation}\label{Eq7}
\Lambda \left( {\Phi ,\Psi } \right) = \left\{ {\lambda  \in \mathbb{C}:\delta \left( {\lambda ,\Phi } \right) = 0,\delta \left( {\lambda ,\Psi } \right) \ge 0} \right\},
\end{equation}
where
\begin{equation*}
\delta \left( {\lambda, \Phi } \right) = {\left[ {\begin{array}{*{20}{c}}
\lambda \\
I
\end{array}} \right]^*}\Phi \left[ {\begin{array}{*{20}{c}}
\lambda \\
I
\end{array}} \right]
\end{equation*}
\end{definition}
According to \cite{zhu2017FGKYP}, when $\Phi$ and $\Psi$ take different forms, \(\Lambda \) can represent a specific frequency range.
\begin{lemma}\label{Lemma 3}
For the continuous-time setting {fractional order system}, one has
\begin{equation}\label{Eq8}
\Phi  = \left[ {\begin{array}{*{20}{c}}
0&{{\rm{e}^{j\theta  }}}\\
{{\rm{e}^{ - j\theta }}}&0
\end{array}} \right]\\
{\rm{     }},
\end{equation}
\begin{equation}\label{Eq8}
\Lambda = \left\{ {{{\left( {\rm{j}\omega } \right)}^\alpha }:\omega  \in \Omega } \right\},
\end{equation}
where \(\theta  = \frac{\pi }{2}\left( {1 - \alpha } \right)\) and $\Omega$ is a subset of real numbers which is determined by the choice of $\Psi$. For different frequency, we can get a table as follows\\

\hspace{-8pt}\begin{tabular}{|c|c|c|c|}
\hline
 &\hspace{-4pt}Low Frequency\hspace{-10pt}&\hspace{-4pt}Middle Frequency\hspace{-4pt}&\hspace{-4pt}High Frequency\hspace{-4pt}\\
 \hline
 $\hspace{-2pt}\Omega\hspace{-2pt}$
&\(0 \le \omega  \le {\omega _L}\)&\(0 \le {\omega _1} \le \omega  \le {\omega _2}\)&\(0 \le {\omega _{_H}} \le \omega \)\\
  \hline
 $\hspace{-2pt}\Psi\hspace{-2pt}$
&\(\hspace{-6pt}\left[ {\begin{array}{*{20}{c}}
{ - 1}&0\\
0&{\omega _{_L}^{2\alpha }}
\end{array}} \right]\)\hspace{-6pt}
&\(\hspace{-6pt}\left[ {\begin{array}{*{20}{c}}
{ - 1}&{{\omega _c}}\\
{{{\overline \omega  }_c}}&{ - \omega _1^\alpha \omega _2^\alpha }
\end{array}} \right]\hspace{-6pt}\)
&\(\hspace{-6pt}\left[ {\begin{array}{*{20}{c}}
1&0\\
0&{ - \omega _H^{2\alpha }}
\end{array}} \right]\hspace{-6pt}\)
\\
  \hline
\end{tabular}
\vspace{6pt}
\\
where \({\omega _c} = \frac{{{\rm{j}^\alpha }\left( {\omega _1^\alpha  + \omega _2^\alpha } \right)}}{2}\).
{In order to ensure that \(\omega \) belongs to the main Riemann sheet, here \(\omega \) must be nonnegative.} This is determined by the particularity of the fractional order system.
\begin{IEEEproof}
{Substituting matrix $\Phi$  into the definition of \(\delta \left( {\lambda ,\Phi } \right)\), there is
\begin{eqnarray*}
\begin{array}{l}
\delta \left( {\lambda ,\Phi } \right) = {\left[ {\begin{array}{*{20}{c}}
\lambda \\
I
\end{array}} \right]^*}\left[ {\begin{array}{*{20}{c}}
0&{{\rm{e}^{{\rm{j}}\theta }}}\\
{{\rm{e}^{ - {\rm{j}}\theta }}}&0
\end{array}} \right]\left[ {\begin{array}{*{20}{c}}
\lambda \\
I
\end{array}} \right]\\
\ \ \ \ \ \ \ \ \ \ = \rm{ cos}\theta {\rm{Re}}\left( \lambda  \right) + \sin \theta {\mathop{\rm Im}\nolimits} \left( \lambda  \right)\\
  \ \ \ \ \ \ \ \ \ \ = 0
\end{array}
\end{eqnarray*}
Therefore, matrix $\Phi$ represents that the object considered here is a continuous-time system.} When
\begin{eqnarray*}
\Psi  = \left[ {\begin{array}{*{20}{c}}
{ - 1}&0\\
0&{\omega _L^{2\alpha }}
\end{array}} \right],
\end{eqnarray*}
there is
 \begin{eqnarray*}
\left( {\lambda {\rm{ + }}\omega _L^\alpha } \right)\left( {\lambda  - \omega _L^\alpha } \right) \le 0 \Rightarrow \lambda  \le \omega _L^\alpha.
\end{eqnarray*}
Hence, $\Lambda$ represents the low frequency. \\
When
\begin{equation*}
\Psi  = \left[ {\begin{array}{*{20}{c}}
1&0\\
0&{{\rm{ - }}\omega _H^{2\alpha }}
\end{array}} \right],
\end{equation*}
there is
\begin{eqnarray*}
\left( {\lambda {\rm{ + }}\omega _H^\alpha } \right)\left( {\lambda  - \omega _H^\alpha } \right) \geq 0 \Rightarrow \lambda  \geq \omega _H^\alpha.
\end{eqnarray*}
Hence, $\Lambda$ represents the high frequency.
As for the middle frequency situation,
there is
\begin{eqnarray*}
\left( {\lambda {\rm{ - }}\omega _1^\alpha } \right)\left( {\lambda  - \omega _2^\alpha } \right) \leq 0 \Rightarrow \omega _1^\alpha \leq \lambda  \leq \omega _2^\alpha.
\end{eqnarray*}
It is worth noting that here, $\omega_c$ is used to guarantee the matrix $\Psi  \in {\mathbb{H}_2}$.
Since $\lambda  = {\omega ^\alpha }$, the conclusion is reached.
\end{IEEEproof}

\begin{remark}\label{Remark 2}
Note that when \(\Psi  \ge 0\), \(\Omega \) can represent the infinite frequency range.
If \(\Psi = {\mathbf{0}_2}\), there is no constraint on $\lambda$, so it can represent the full band.  If  $\Psi  > 0$, depending on the property of positive definite matrix, for any non-zero vector $X$, there is \({X^*}\Psi X > 0\). Let \(X = \left[\hspace{-4pt} {\begin{array}{*{20}{c}}
\lambda \\
{{I}}
\end{array}} \hspace{-4pt}\right]\), {and one can then find that there is no constraint on $\lambda$ by simple calculation, so \(\Omega \) can also represent} {the full frequency range} when \(\Psi  > 0\). In this study, \(\Psi = {\mathbf{0}_2}\) is simply taken.
\end{remark}
\end{lemma}

\section{GKYP lemma for SFOS}\label{Section 3}
{In this section, the appropriate $\mathbb{S}_1$ and $\mathbb{M}$ will be chosen to derive the GKYP lemma for SFOFs by the S-procedure tool, and the conclusion will be strictly proved.}

According to the standard KYP lemma in \cite{Rantzer1996KYP-lemma} and the transfer function of SFOS in (2), the set $\mathbb{S}_1$ which represents the positive definiteness of the SFOS should be given as
\begin{equation}\label{Eq9}
\hspace{0pt}{\mathbb{S}_1} = \hspace{-3pt}\left\{ {\xi {\xi ^*}\hspace{-5pt}:\hspace{0pt}\xi  = \hspace{-3pt}\left[\hspace{-6pt} {\begin{array}{*{20}{c}}
{{{\left[ {{{\left( {\rm{j}\omega } \right)}^\alpha }E - A} \right]}^{ - 1}}B}\\
I
\end{array}} \hspace{-6pt}\right]\eta ,\hspace{-3pt}\begin{array}{*{20}{c}}
{\eta  \in {C^m},\eta  \ne 0}\\
{\omega  \in {\mathbb{R}^ + }\hspace{-3pt} \cup \left\{ +\infty  \right\}}
\end{array}} \hspace{-6pt}\right\}.\hspace{-8pt}
\end{equation}
This set can be described as
\begin{equation}\label{Eq10}
\begin{array}{l}
{\mathbb{S}_1} = \left\{ {\xi {\xi ^*}:\xi  \in {\mathbb{G}_\lambda },\lambda  \in \overline \Lambda  } \right\},\\
{\mathbb{G}_\lambda } = \left\{ {\xi  \in {\mathbb{C}^{n + m}}:\xi  \ne 0,{\Gamma _\lambda }F\xi  = 0} \right\},
\end{array}
\end{equation}
where $\overline \Lambda   = {\left( {\rm{j}{\mathbb{R}^ + }} \right)^\alpha } \cup \left\{ \infty  \right\} $ and
\begin{eqnarray}\label{Eq11}
{\Gamma _\lambda } = \left\{ {\begin{array}{*{20}{c}}
{[\begin{array}{*{20}{c}}
{{I_n}}&{ - \lambda {I_n}}
\end{array}]\ \ {\rm{   (}}\lambda  \in \mathbb{C}){\rm{ }}}\\
{{\rm{  }}\left[ {\begin{array}{*{20}{c}}
0&{ - {I_n}}
\end{array}} \right]\ \ {\rm{    (}}\lambda  = \infty {\rm{)}}}
\end{array}} \right., {\ F = \left[ {\begin{array}{*{20}{c}}
A&B\\
{{E}}&0
\end{array}} \right]}.
\end{eqnarray}
{Considering} the general frequency range $\Lambda$ in (6), the $\overline \Lambda$ in (10) is defined as
\begin{equation}\label{Eq12}
\overline \Lambda   = \left\{ {\begin{array}{*{20}{c}}
{\Lambda ,\qquad \ \ \text{if}  \ \Lambda \  \text{is bounded}};\\
{\Lambda  \cup \left\{ \infty  \right\},\qquad \text{otherwise}}.
\end{array}} \right.
\end{equation}
Now, the main technical steps in achieving the GKYP lemma for the SFOS are to represent the set $\mathbb{S}_1$ in (10) as (4) through choosing a suitable $\mathbb{M}$.  At the same time, in order to ensure that the S-procedure is lossless, it is also necessary to indicate that the selected set $\mathbb{M}$ possesses the attributes in Definition 1.
\begin{lemma}\label{Lemma 4} \cite{zhu2017FGKYP}
Let \({\Phi _0},{\Psi _0} \in {\mathbb{H}_2}\)， and a nonsingular matrix \(T \in {\mathbb{C}^{2 \times 2}}\) be given and define \(\Phi ,\Psi  \in {\mathbb{H}_2}\) as follows
\begin{equation}\label{Eq13}
\begin{array}{l}
\Phi  = {T^*}{\Phi _0}T,\ \
\Psi  = {T^*}{\Psi _0}T,
\end{array}
\end{equation}
\begin{equation}\label{Eq14}
\begin{array}{l}
{\Phi _0} = \left[ {\begin{array}{*{20}{c}}
0&{{\rm{e}^{\rm{j}\theta  }}}\\
{{\rm{e}^{ - \rm{j}\theta  }}}&0
\end{array}} \right],\ \
{\Psi _0} = \left[ {\begin{array}{*{20}{c}}
\alpha &{\beta {\rm{e}^{\rm{j}\theta  }}}\\
{\beta {\rm{e}^{ - \rm{j}\theta  }}}&\gamma
\end{array}} \right],
\end{array}
\end{equation}
where \(\alpha ,\beta ,\gamma  \in \mathbb{R}\) ,\(\alpha  \le \gamma\), and \( \gamma  \ge 0\).
Consider \({\Gamma _\lambda }\) in (11), \(\Lambda \) in (6) and \(\overline \Lambda  \) in (12). Suppose \(\overline \Lambda  \) represents curves. For a given vector \(\zeta  \in {\mathbb{C}^{2n}}\), the following two conditions are equivalent.\\
i) \({\Gamma _\lambda }\zeta  = 0\) holds for some \(\lambda  \in \overline \Lambda  \left( {\Phi ,\Psi } \right)\).\\
ii) \({\Gamma _s}\left( {T \otimes I} \right)\zeta  = 0\) holds for some \(s \in \overline \Lambda  \left( {{\Phi _0},{\Psi _0}} \right)\).
\end{lemma}
\begin{lemma}\label{Lemma 5} \cite{zhu2017FGKYP}
Let \({{\Phi _0},{\Psi _0}}\) be defined in (14), \({\Gamma _\lambda }\) in (11), \(\Lambda \) in (6) and \(\overline \Lambda  \) in (12). Suppose \(\overline \Lambda  \) represents curves. The following conditions are equivalent.\\
i) \({\Gamma _s}\eta  = 0\) for some \(s \in \overline \Lambda  \left( {{\Phi _0},{\Psi _0}} \right)\);\\
ii) \({\eta ^*}\left( {{\Phi _0} \otimes P + {\Psi _0} \otimes Q} \right)\eta  \ge 0\) for all \(P,Q \in {\mathbb{H}_n},Q \ge 0\).\\
\end{lemma}

\begin{theorem}
Let \(F \in {\mathbb{C}^{2n \times (n + m)}}\) and define \({\Gamma _\lambda }\) and \(\overline \Lambda  \) as (11) and (12). The matrices \(\Phi ,\Psi  \in {\mathbb{H}_2}\) are given such that $\Lambda $ in (6) represents curves. Then the set \(\mathbb{S}_1\) defined in (10) can be represented by (5) and (6) with
\begin{equation}\label{Eq15}
\mathbb{M} = \{ {F^*}(\Phi  \otimes P + \Psi  \otimes Q)F:P,Q \in {\mathbb{H}_n},Q \ge 0\}.
\end{equation}
\end{theorem}

\begin{IEEEproof}
Let $\mathbb{S}_1$ be defined by (10) and $\mathbb{S}_2$ be defined to be ${\mathbb{S}_1}$ in (5) with $\mathbb{M}$ in (15). \({F_0} = \left( {T \otimes I} \right)F\). Then, for a nonzero vector \(\xi \), one has
\begin{equation}\label{Eq16}
\begin{array}{l}
\quad\ \xi {\xi ^*} \in {\mathbb{S}_1}\\
 \Leftrightarrow {\Gamma _\lambda }F\xi  = 0\ \text{for some}\ \lambda  \in \overline \Lambda  \left( {\Phi ,\Psi } \right)\\
 \Leftrightarrow {\Gamma _s}{F_0}\xi  = 0\ \text{for some}\ s \in \overline \Lambda  \left( {{\Phi _0},{\Psi _0}} \right)\\
 \Leftrightarrow {\xi ^*}F_0^*\left( {{\Phi _0} \otimes P + {\Psi _0} \otimes Q} \right)\xi {F_0} \ge 0\\
\quad\ \text{for all}\ P,Q \in {\mathbb{H}_n},Q \ge 0\\
 \Leftrightarrow \xi {\xi ^*} \in {\mathbb{S}_2},
\end{array}
\end{equation}
where the first and fourth equivalences can be derived from the definition, the second equivalence holds based on  Lemma 4, and the third equivalence holds due to Lemma 5.
\end{IEEEproof}
Note that for a singular fractional order system, the matrix $F$ that affects the set $\mathbb{M}$ is different from the normal system , so we need to specifically prove that $\mathbb{M}$ satisfies Definition 1.
\begin{theorem}
 Let {\(F = \left[ {\begin{array}{*{20}{c}}
A&B\\
{{E}}&0
\end{array}} \right]\) }and \(\Phi ,\Psi  \in {\mathbb{H}_2}\) be given such that $\Lambda $ in (6) represents curves. Define \({\Gamma _\lambda }\) by (11) , \(\overline \Lambda  \) by (12), and the set $\mathbb{M}$ by (15). Then the set $\mathbb{M}$ is admissible and rank-one separable.
\end{theorem}
\begin{IEEEproof}The proof process is divided into the following two steps.\\
\textbf{Step 1.} {The set $\mathbb{M}$ is a closed convex cone by definition.} When \(M \in \mathbb{M} > 0\), the set $\mathbb{S}_1$ is not empty. According to Lemma 11 in \cite{Iwasaki2005GKYPlemma}, $\mathbb{M}$ is admissible.\\
\textbf{Step 2.} According to Lemma 4, we get
\begin{eqnarray}\label{Eq17}
\begin{array}{l}
\Phi  \otimes P + \Psi  \otimes Q\ \ \ \ \ \\
 \hspace{-10pt}=\hspace{-3pt} {(T \otimes I)^*} \hspace{-3pt}\left[ \hspace{-3pt}{\begin{array}{*{20}{c}}
{\alpha Q}&{\hspace{-10pt}P{\rm{e}^{\rm{j}\theta }} + \beta {\rm{e}^{\rm{j}\theta }}Q}\\
{P{\rm{e}^{ - \rm{j}\theta }} + \beta {\rm{e}^{ - \rm{j}\theta }}Q}&{\hspace{-10pt}\gamma Q}
\end{array}} \hspace{-3pt}\right] \hspace{-3pt} \left( {T \otimes I} \right),
\end{array}
\end{eqnarray}
where \(\alpha  < 0 < \gamma \) or \(\gamma  \ge \alpha  \ge 0\).

When \(\gamma  \ge \alpha  \ge 0\), let
{\begin{eqnarray}\label{Eq21}
\hspace{3pt} V = \left[ {\begin{array}{*{20}{c}}
{{\rm{e}^{ - \rm{j}\theta /2}}}&0\\
0&{{\rm{e}^{\rm{j}\theta /2}}}
\end{array}} \right]\left( {T \otimes I} \right)\left[ {\begin{array}{*{20}{c}}
A&B\\
{{E}}&0
\end{array}} \right],\ \ \ \ \ \ \ \ \ \
\end{eqnarray}}
{\begin{eqnarray}
X = P + \beta Q,\ \ \ \ \ \ \ \ \ \ \ \ \ \ \ \ \ \ \ \ \ \ \ \ \ \ \ \ \ \ \ \ \ \ \ \ \ \ \ \ \ \ \ \  \ \
\end{eqnarray}
\begin{eqnarray}\label{Eq22}
\begin{array}{l}
Y = {\left[ {\left( {T \otimes I} \right)F} \right]^*}\left[ {\begin{array}{*{20}{c}}
{\alpha Q}&0\\
0&{\gamma Q}
\end{array}} \right]\left( {T \otimes I} \right)\left[ {\begin{array}{*{20}{c}}
A&B\\
{{E}}&0
\end{array}} \right].
\end{array}
\end{eqnarray}}
Then, the set $\mathbb{M}$ can be expressed as \(\mathbb{M} = {V^*}{\mathbb{M}_X}V + Y\) with $\mathbb{M}_X$ defined as
\begin{equation}\label{Eq23}
{\mathbb{M}_X} = \left\{ {\left[ {\begin{array}{*{20}{c}}
0&X\\
X&0
\end{array}} \right]:X \in {\mathbb{H}_n}} \right\}.
\end{equation}

When \(\alpha  < 0 < \gamma \), let
{\begin{equation}\label{Eq18}
W = \left[ {\begin{array}{*{20}{c}}
{\sqrt { - \alpha } I{\rm{e}^{ - j\varphi /2}}}&\hspace{-16pt}0\\
0&\hspace{-16pt}{\sqrt \gamma  I{\rm{e}^{j\varphi /2}}}
\end{array}} \right](T \otimes I)\left[ {\begin{array}{*{20}{c}}
A&B\\
{{E}}&0
\end{array}} \right],
\end{equation}}
\begin{equation}\label{Eq19}
X = \frac{{(P + \beta Q)}}{{\sqrt { - \alpha \gamma } }},\ Y = Q.
\end{equation}
Then, the set $\mathbb{M}$ can be expressed as
\(\mathbb{M} = {W^*}{\mathbb{M}_{XY}}W\) with \({\mathbb{M}_{XY}}\) defined as
\begin{equation}\label{Eq20}
{\mathbb{M}_{XY}} = \left\{ {
\begin{smallmatrix}
\left[ {\begin{array}{*{20}{c}}
{ - Y}&X\\
X&Y
\end{array}} \right]\end{smallmatrix}:X,Y \in {\mathbb{H}_n},Y \ge 0} \right\}.
\end{equation}

Since $\mathbb{M}_X$ and $\mathbb{M}_{XY}$ are proved rank-one separable in \cite{Iwasaki2005GKYPlemma}, it then follows from Lemma 2 that $\mathbb{M}$ is rank-one separable. This ends the proof.
\end{IEEEproof}

\begin{theorem}\label{Theorem 3}(GKYP lemma for SFOS)
Let A \(\in {\mathbb{R}^{n \times n}},B \in {\mathbb{R}^{n \times m}},\Theta  \in {\mathbb{H}_{n + m}}\) and \(\Phi ,\Psi  \in {\mathbb{H}_2}\) be given. Define \(\Lambda \ and\ \overline \Lambda  \) by (6) and (12). \({\Gamma _\lambda }\) is defined in (11) and \({S_\lambda }\) is defined as the null space of \({\Gamma _\lambda }F\). Suppose $\Lambda$ represents curves on the right half complex plane, the following statements are equivalent\\
i)\ \(S_\lambda ^*\Theta {S_\lambda } < 0,\  \forall \lambda  \in \overline \Lambda  \left( {\Phi ,\Psi } \right)\).\\
ii)\ There exist \(P,Q \in {\mathbb{H}_n}\) such that \(Q > 0\) and
\begin{equation}\label{Eq24}
{\left[ {\begin{array}{*{20}{c}}
A&B\\
{{E}}&0
\end{array}} \right]^*}\left( {\Phi  \otimes P + \Psi  \otimes Q} \right)\left[ {\begin{array}{*{20}{c}}
A&B\\
{{E}}&0
\end{array}} \right] + \Theta  < 0.
   \end{equation}
\end{theorem}
\begin{IEEEproof}
i) holds if and only if \({\rm{tr}}(\Theta {\mathbb{S}_1}) < 0\) where \({\mathbb{S}_1}\) is defined in (10) with $\mathbb{M}$ in (15). By Theorem 2, the set $\mathbb{M}$ is admissible and rank-one separable. Hence, according to (5), condition i) is equivalent to \(\Theta  +  \mathbb{M} < 0\). Substituting the set $\mathbb{M}$  in (15) into the above inequality, it is concluded that (25) holds when there exist \(P,Q \in {\mathbb{H}_n}\) such that \(Q \ge 0\). Since the inequality in (25) is strict, the positivity of $Q$ can be enhanced to $Q > 0$ without loss of generality.
\end{IEEEproof}

\section{\({L_\infty }\) bounded real lemmas for the SFOS}\label{Section 4}
{In this section, the \({L_\infty }\) bound real lemmas for the SFOS in different frequency ranges will be derived.}

{As is known, for a matrix singular \(G(s)\), the \({L_\infty }\) norm of \(G(s)\) is defined as
\({\left\| {G\left( {\rm{s}} \right)} \right\|_{{L_\infty }}} = \mathop {\sup }\limits_{\omega  \in \mathbb{R}} {\sigma _{\max }}\left( {G(\rm{j}\omega )} \right),
\)
where \({\sigma _{\max }}\) is the maximum singular value. It has the following property.}

\begin{lemma}\label{Lemma8}\cite{liang2015bounded}
For a matrix function G(s), there holds
\begin{eqnarray}\label{Eq26}
{\left\| {G\left( {\rm{s}} \right)} \right\|_{{L_\infty }}} = \mathop {\sup }\limits_{\omega  \ge 0} {\sigma _{\max }}\left( {G(\rm{j}\omega )} \right),
\end{eqnarray}
\end{lemma}

\begin{theorem}\label{Theorem 4}(L-BR Lemma for SFOSs at Low Frequency)
Consider an SFOS whose transfer function G(s) is (2). {If the \({L_\infty }\) performance bound is given as \(\delta  > 0\), then for all
\(\omega \) belong to the principal Riemann sheet and \(\omega  \in {\Omega _L} = \left\{ {\omega  \in {\mathbb{R}^ + }:0 \le \omega  \le {\omega _L}} \right\}\), \({\left\| {G\left( {\rm{s}} \right)} \right\|_{{L_\infty }}}< \delta\) holds if and only if there exist \(P,Q \in {\mathbb{H}_{n}},Q > 0\), such that}
\begin{eqnarray}\label{Eq27}
\left[ {\begin{array}{*{20}{c}}
{{\rm {sym}}({X{E}}) - {A^{\rm{T}}}QA + W}&{{Y^*}}&{{C^{\rm{T}}}}\\
Y&{ \hspace{-20pt}- \delta I - {B^{\rm{T}}}QB}&{{D^{\rm{T}}}}\\
C&D&{ - \delta I}
\end{array}} \right] < 0,
\end{eqnarray}
where \({X = {{\rm{e}}^{\rm{j}\theta }}{A^{\rm{T}}}P},\ Y =  - {B^{\rm{T}}}QA + {{\rm{e}}^{\rm{j}\theta }}{B^{\rm{T}}}P{E}\) , \(W = {E^{\rm{T}}\omega _L^{2\alpha }Q{E}}\), \(\theta  = \frac{\pi }{2}\left( {1 - \alpha } \right)\).
\end{theorem}
\begin{IEEEproof}
For low frequency, according to  Lemma 3, let
 \begin{eqnarray*}
 \Phi  = \left[ {\begin{array}{*{20}{c}}
0&{{\rm{e}^{\rm{j}\theta }}}\\
{{\rm{e}^{ - \rm{j}\theta }}}&0
\end{array}}\right],\Psi  = \left[ {\begin{array}{*{20}{c}}
{ - 1}&0\\
0&{\omega _L^{2\alpha }}
\end{array}} \right].
\end{eqnarray*}
 Then, according to Lemma 4, \(\overline \Lambda  \left( {\Phi ,\Psi } \right)\) can represent a curve on complex plane with the frequency range \({\Omega _L}\).

Let \(\lambda \left( \omega  \right) = {\rm{e}^{\rm{j}\frac{\pi }{2}\alpha }}{\omega ^\alpha }\), and then
\begin{eqnarray*}
G\left( {{\rm{j}}\omega } \right) = C{\left[ {\lambda \left( \omega  \right){E} - A} \right]^{ - 1}}B + D.
\end{eqnarray*}

According to the definition of \({\sigma _{\max }}\), one has
\begin{equation}\label{Eq28}
\begin{array}{l}
\ \ \ \ \mathop {\sup }\limits_{\omega  \in \mathbb{R}} {\sigma _{\max }}\left( {G\left( {\rm{j}\omega } \right)} \right) < \delta \\
 \Leftrightarrow {G^*}\left( {\rm{j}\omega } \right)G\left( {\rm{j}\omega } \right) - {\delta ^2}I < 0,\forall \omega  \in {\Omega _L}\\
 \Leftrightarrow {\left[ {\begin{array}{*{20}{c}}
{H\left( \lambda  \right)}\\
I
\end{array}} \right]^*}\theta \left[ {\begin{array}{*{20}{c}}
{H\left( \lambda  \right)}\\
I
\end{array}} \right] < 0,\forall \lambda  \in \overline \Lambda  \left( {\Phi ,\Psi } \right).
\end{array}
\end{equation}

where
\begin{equation*}
H\left( \lambda  \right) = {\left( {\lambda {E} - A} \right)^{ - 1}}B,\
\theta  = \left[ {\begin{array}{*{20}{c}}
{{C^{\rm{T}}}C}&{{C^{\rm{T}}}D}\\
{{D^{\rm{T}}}C}&{{D^{\rm{T}}}D - {\delta ^2}I}
\end{array}} \right].
\end{equation*}

According to Theorem 3, the last part of (28) is equivalent to the following LMI with \(P,Q \in {\mathbb{H}_{\rm{n}}},Q > 0\).
\begin{equation}\label{Eq31}
{\left[ {\begin{array}{*{20}{c}}
A&B\\
{{E}}&0
\end{array}} \right]^*}\left( {\Phi  \otimes P + \Psi  \otimes Q} \right)\left[ {\begin{array}{*{20}{c}}
A&B\\
{{E}}&0
\end{array}} \right] + \theta  < 0.
\end{equation}
Substituting the definitions of \(\Phi \ {\rm{and}}\ \Psi\) into the above equation, it follows
\begin{eqnarray}\label{Eq32}
\begin{array}{l}
{\left[ {\begin{array}{*{20}{c}}
A&B\\
{{E}}&0
\end{array}} \right]^*} {\left[ {\begin{array}{*{20}{c}}
{ - Q}&{{\rm{e}^{\rm{j}\theta }}P}\\
{{\rm{e}^{ - \rm{j}\theta }}P}&{{\omega ^{2\alpha }}Q}
\end{array}} \right]} \left[ {\begin{array}{*{20}{c}}
A&B\\
{{E}}&0
\end{array}} \right]+ \theta  < {0}.
\end{array}
\end{eqnarray}
Let \({X = {{\rm{e}}^{\rm{j}\theta }}{A^{\rm{T}}}P}\) , \( Y =  - {B^{\rm{T}}}QA + {{\rm{e}}^{\rm{j}\theta }}{B^{\rm{T}}}P{E}\), \(W = {E^{\rm{T}}\omega _L^{2\alpha }Q{E}}\) and \({B_{\delta}} =  - {\delta ^2}I - {B^{\rm T}}QB\). The LMI (30) can be simplified as
\begin{equation}\label{Eq33}
\left[ \hspace{-4pt} {\begin{array}{*{20}{c}}
{{\rm{sym}}\left({ X{E} }\right)- {A^{\rm{T}}}QA + W} \hspace{-3pt}&{{Y^*}}\\
Y\hspace{-10pt}&{ B_{\delta}}
\end{array}}\hspace{-4pt} \right] \hspace{-3pt} +\hspace{-3pt}  \left[\hspace{-5pt} {\begin{array}{*{20}{c}}
{{C^{\rm{T}}}}\\
{{D^{\rm{T}}}}
\end{array}} \hspace{-5pt}\right] \left[ \hspace{-5pt} {\begin{array}{*{20}{c}}
C&\hspace{-5pt}D
\end{array}}  \hspace{-5pt}\right] < 0.
\end{equation}
Then according to the Schur complement theorem in \cite{zhang2006schur}, LMI (27) is finally achieved. This completes the proof.
\end{IEEEproof}

\begin{theorem}\label{Theorem 5}
(L-BR Lemma for SFOSs at Middle Frequency)
Consider an SFOS whose transfer function G(s) is (2). {If \({L_\infty }\) performance bound is given as \(\delta  > 0\), then for all
\(\omega \) belong to the principal Riemann sheet and \(\omega  \in {\Omega _M} = \left\{ {\omega  \in {\mathbb{R}^ + }:0 < {\omega _1} < \omega  < {\omega _2}} \right\}\), \({\left\| {G\left( {\rm{s}} \right)} \right\|_{{L_\infty }}}  < \delta\) holds if and only if there exist \(P,Q \in {\mathbb{H}_{n}},Q > 0\), such that}
\hspace{-10pt}\begin{eqnarray}\label{Eq34}
\left[ {\begin{array}{*{20}{c}}
{{\rm {sym}}({X{E}}) - {A^{\rm{T}}}QA - W}&{{Y^*}}&{{C^{\rm{T}}}}\\
Y&{\hspace{-20pt} - \delta I - {B^{\rm{T}}}QB}&{{D^{\rm{T}}}}\\
C&D&{ - \delta I}
\end{array}} \right] < 0,
\end{eqnarray}
where \({X = {{\rm{e}}^{j\theta }}{A^{\rm{T}}}P},\ Y =  - {B^{\rm{T}}}QA + {{\rm{e}}^{j\theta }}{B^{\rm{T}}}P{E}\),  \(W = E^{\rm{T}}\omega _1^\alpha \omega _2^\alpha Q{E}\), \(\theta  = \frac{\pi }{2}\left( {1 - \alpha } \right)\) .
\end{theorem}

\begin{IEEEproof}
The theorem of middle frequency can be proved similar to the proof of low frequency. According to Lemma 3, the curve \(\overline \Lambda  \left( {\Phi ,\Psi } \right)\) here is chosen as
\begin{eqnarray}\label{Eq37}
\Phi  = \left[ {\begin{array}{*{20}{c}}
0&{{\rm{e}^{\rm{j}\theta }}}\\
{{\rm{e}^{ - \rm{j}\theta }}}&0
\end{array}} \right],\Psi  = \left[\hspace{-4pt}{\begin{array}{*{20}{c}}
{ - 1}&{{\rm{j}^\alpha }\frac{{\omega _1^\alpha  + \omega _2^\alpha }}{2}}\\
{{{\left(  { - \rm{j}} \right)}^\alpha }\frac{{\omega _1^\alpha  + \omega _2^\alpha }}{2}}&{ - \omega _1^\alpha \omega _2^\alpha }
\end{array}}  \hspace{-4pt}\right].
\end{eqnarray}
Then following the proof of Theorem 4, Theorem 5 can be proved. To avoid duplication, the remaining proof process is omitted here.
\end{IEEEproof}
\begin{theorem}\label{Theorem 6}
(L-BR Lemma for SFOSs at High Frequency)
Consider an SFOS whose transfer function G(s) is (2). {If \({L_\infty }\) performance bound is given as \(\delta  > 0\), then for all
\(\omega \) belong to the principal Riemann sheet and \(\omega  \in {\Omega _H} = \left\{ {\omega  \in {\mathbb{R}^ + }:0 \le \omega _H \le {\omega }} \right\}\), \({\left\| {G\left( {\rm{s}} \right)} \right\|_{{L_\infty }}} < \delta\) holds if and only if there exist \(P,Q \in {\mathbb{H}_{n}},Q > 0\), such that}
\begin{eqnarray}\label{Eq35}
\left[ {\begin{array}{*{20}{c}}
{{\rm{sym}}({X{E}}) + {A^{\rm{T}}}QA - W}&{{Y^*}}&{{C^{\rm{T}}}}\\
Y&{ \hspace{-20pt}- \delta I + {B^{\rm{T}}}QB}&{{D^{\rm{T}}}}\\
C&D&{ - \delta I}
\end{array}} \right] < 0,
\end{eqnarray}
where \({X = {{\rm{e}}^{\rm{j}\theta }}{A^{\rm{T}}}P},\ Y =  - {B^{\rm{T}}}QA + {{\rm{e}}^{\rm{j}\theta }}{B^{\rm{T}}}P{E}\) , \(W = {E^{\rm{T}}\omega _H^{2\alpha }Q{E}}\), \(\theta  = \frac{\pi }{2}\left( {1 - \alpha } \right)\) .
\end{theorem}
\begin{IEEEproof}
The curve \(\overline \Lambda  \left( {\Phi ,\Psi } \right)\) in high frequency is chosen as
\begin{eqnarray}\label{Eq38}
\Phi  = \left[ {\begin{array}{*{20}{c}}
0&{{\rm{e}^{\rm{j}\theta }}}\\
{{\rm{e}^{ - \rm{j}\theta }}}&0
\end{array}} \right],\Psi  = \left[ {\begin{array}{*{20}{c}}
1&0\\
0&{ - \omega _H^{2\alpha }}
\end{array}} \right].
\end{eqnarray}
Similar to the proof process of the previous two theorems, Theorem 6 can be proved to be true.
\end{IEEEproof}
\begin{theorem}\label{Theorem 7}
(L-BR Lemma for SFOSs at Full Frequency)
Consider an SFOS whose transfer function G(s) is (2). Given a prescribed \({L_\infty }\) performance bound  \(\delta  > 0\), then
\({\left\| {G\left( s\right)} \right\|_{{L_\infty }}} = \mathop {\sup }\limits_{\omega  \in \mathbb{R}} {\sigma _{\max }}\left( {G(\rm{j}\omega )} \right) < \delta\), where \(\omega \) belongs to the principal Riemann sheet and \(\Omega  \in {\Omega _0} = {\mathbb{R}^ + } \cup \left\{ { + \infty } \right\}\), holds if and only if there exists \(P\in {\mathbb{H}_{n}}\), such that
\begin{eqnarray}\label{Eq35}
\left[ {\begin{array}{*{20}{c}}
{{\rm{sym}}(XA)}&{{XB}}&{{C^{\rm{T}}}}\\
(XB)^*&{ - \delta I}&{{D^{\rm{T}}}}\\
C&D&{ - \delta I}
\end{array}} \right] < 0,
\end{eqnarray}
where \(X = {\rm{e}^{ - \rm{j}\theta  }}E^{\rm{T}}P^*\), \(\theta  = \frac{\pi }{2}\left( {1 - \alpha } \right)\) .
\end{theorem}
\begin{IEEEproof}
The curve \(\overline \Lambda  \left( {\Phi ,\Psi } \right)\) in infinite frequency is chosen as
\begin{eqnarray}\label{Eq38}
\Phi  = \left[ {\begin{array}{*{20}{c}}
0&{{\rm{e}^{\rm{j}\theta  }}}\\
{{\rm{e}^{ - \rm{j}\theta  }}}&0
\end{array}} \right],\Psi  = \left[ {\begin{array}{*{20}{c}}
0&0\\
0&0
\end{array}} \right].
\end{eqnarray}
Using these two matrices for calculation, Theorem 7 can be proved to be correct.
\end{IEEEproof}
\begin{remark}
This is the first time to establish the universal framework of the GKYP lemma for the SFOS. When the order \(\alpha  = 1\), it degenerates into the GKYP lemma of the integer order singular system. Therefore, the conclusion here is more general and universal. {Besides, for singular systems, the GKYP lemma established here is completely new and the conclusion is proven systematically for the first time, regardless of fractional or integer order systems.}
\end{remark}
\section{\({L_\infty }\) controller synthesis for the SFOS}\label{Section 5}
As the \({L_\infty }\) performance index of the system matrix is very important to a system, we can design the controller to modify the system which does not meet our requirements.

Considering the SFOS in (1), state feedback will be used to design the controller. Let \(u\left( t \right) = v\left( t \right)  + Kx\left( t \right)\), where \(v\left( t \right)  \in {\mathbb{R}^n}\) is the exogenous input. Then the closed loop system has the transfer function from $v$ to $y$ as
\begin{eqnarray}
{G_{vy}}\left( s \right) = \left( {C + DK} \right){\left[ {{s^\alpha }{E} - \left( {A + BK} \right)} \right]^{ - 1}}B + D.
\end{eqnarray}

\begin{theorem}
Consider the SFOS system (1) with transfer function ${G_{vy}}\left( s \right) $ in (38). If and only if there exist \(P \in {\mathbb{R}^{n \times n}},Q \in {\mathbb{R}^{m \times n}}\), such that for \(X = {{\rm{e}}^{ \rm{j}\theta }}P,Y = {{\rm{e}}^{ \rm{j}\theta }}Q\) and \(\theta  = \frac{\pi }{2}\left( {1 - \alpha } \right)\), the following LMI is established
\begin{eqnarray}
\left[ \hspace{-4pt}{\begin{array}{*{20}{c}}
{{\rm{sym}}(AXE + BYE)}&\hspace{-6pt}{{{E^{\rm{T}}(CX + DY)}^*}}&\hspace{-6pt}B\\
{(CX + DY)E}&{ - \delta I}&\hspace{-6pt}D\\
{{B^{\rm{T}}}}&{{D^{\rm{T}}}}&{ \hspace{-6pt}- \delta I}
\end{array}} \hspace{-4pt}\right] < 0,
\end{eqnarray}
there holds \({\left\| {{G_{vy}}} \right\|_{{L_\infty }}} < \delta \). The state feedback controller can be derived as \(K = Y{X^{ - 1}}\).
\end{theorem}
\begin{IEEEproof}
In order to facilitate the design of the controller, we should make some mathematical deformation for Theorem 7. Considering the duality of the system, let \(A_1 = {A}^{\rm{T}},\ B_1 = C^{\rm{T}},\ C_1 = B^{\rm{T}}\  ,D_1 = D^{\rm{T}}\), and \(X_1 = {X}^*={\rm{e}^{\rm{j}\theta }}P{E}\), and then (31) can be replaced by
\begin{eqnarray}
\begin{array}{l}
{\left[\hspace{-4pt} {\begin{array}{*{20}{c}}
{{A^*}}&{{C^*}}\\
{{E}}&0
\end{array}} \hspace{-4pt}\right]^*}\left( {\Phi  \otimes P + \Psi  \otimes Q} \right)\left[ \hspace{-4pt} {\begin{array}{*{20}{c}}
{{A^*}}&{C^*}\\
{{E}}&0
\end{array}}  \hspace{-4pt}\right] + \theta  < 0,
\end{array}
\end{eqnarray}
where
$\theta  = \left[ {\begin{array}{*{20}{c}}
{{C^{\rm{T}}}C}&{{C^{\rm{T}}}D}\\
{{D^{\rm{T}}}C}&{{D^{\rm{T}}}D - {\delta ^2}I}
\end{array}} \right].$

By the derivation step by step, the LMI (36) in Theorem 7 is transformed into
\begin{eqnarray}
\left[ {\begin{array}{*{20}{c}}
{{\rm{sym}}(A_1X_1)}&{{{(C_1X_1)}^*}}&B_1\\
{C_1X_1}&{ - \delta I}&D_1\\
{{B_1^{\rm{T}}}}&{{D_1^{\rm{T}}}}&{ - \delta I}
\end{array}} \right] < 0.
\end{eqnarray}
By comparing $G_{vy}$ in (38) with the original transfer function in (2), there is \(A \Rightarrow A + BK,C \Rightarrow C + DK\). In order to make sure that \(K = Y{X^{ - 1}}\) is real and available, $X$ needs to be nonsingular. Let \({X_1} = {\rm{e}^{\rm{j}\theta }}P{E} = {X_2}{E}\), and then the matrix ${X_2}$ is invertible, and the LMI (39) is reached.

Since the controller gain must be real, the conclusion needs to be further proven. Suppose there exists $\widetilde P \in {\mathbb{H}_n}$ satisfying the LMI (39). According to the property of Hermitian matrix, $\widetilde P ={\widetilde P^*}$, one has
\begin{eqnarray*}
{\rm Re} ( \widetilde P)  = \frac{1}{2}( {\widetilde P + {{\widetilde P}^*}}).
\end{eqnarray*}
Let \(P = {\rm Re} ( \widetilde P) \), and then for each $\widetilde P \in {\mathbb{H}_n}$ satisfying the LMI (39), we can find a corresponding \(P \in {\mathbb{R}^{n \times n}}\) also satisfies the above condition.
\end{IEEEproof}

\section{Simulation Study}\label{Section 6}
\begin{example}
Consider a singular fractional order system with the following state space representation
\begin{eqnarray*}
\left\{ {\begin{array}{*{20}{c}}
{\left[ {\begin{array}{*{20}{c}}
1&0\\
0&0
\end{array}} \right]{\mathscr{D}^{0.5}}x = \left[ {\begin{array}{*{20}{c}}
1&2\\
2&{ - 1}
\end{array}} \right]x + \left[ {\begin{array}{*{20}{c}}
1\\
0
\end{array}} \right]u},\\
\\
{\ \ y = \left[ {\begin{array}{*{20}{c}}
1&1
\end{array}} \right]x.}
\end{array}} \right.
\end{eqnarray*}
Under low frequency conditions, let $0\leq\omega\leq100\ Hz$, the maximum singular values are shown in Fig. 1.

\begin{figure}[!htbp]
\centering
\includegraphics[width=1\columnwidth]{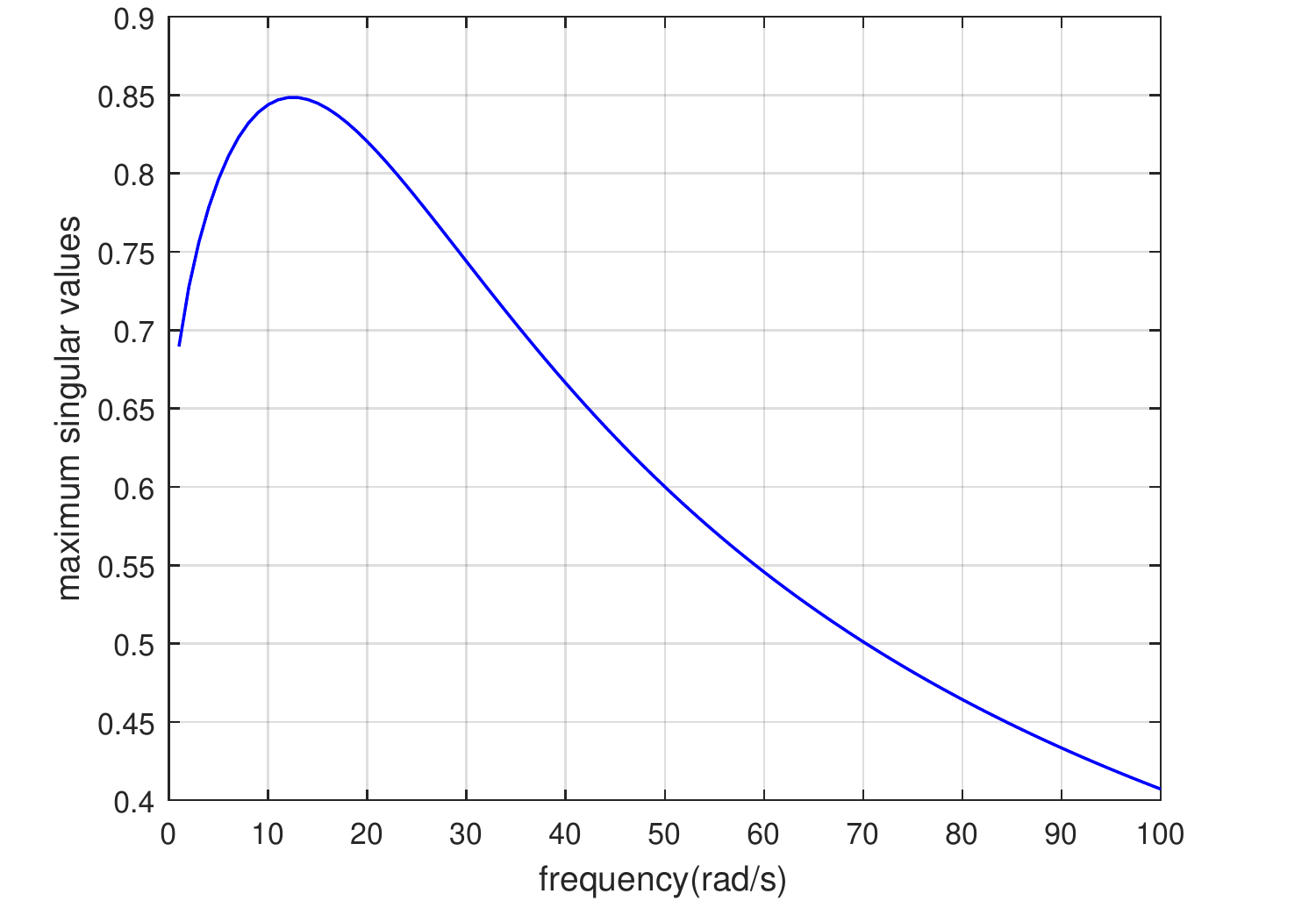}\vspace{-10pt}
\caption{Maximum singular values of Example 1.}\label{Fig1}
\end{figure}
It shows that the \({\left\| {G\left( s \right)} \right\|_{{L_\infty }}}\) index of Example 1 is about 0.85 in the frequency range. Due to Theorem 4, letting \(\delta {\rm{ = }}0.9\), one has
\begin{eqnarray*}
U = \left[ {\begin{array}{*{20}{c}}
{ - 3.1443}&{ - 2.8870}\\
{ - 2.8870}&0
\end{array}} \right],V = \left[ {\begin{array}{*{20}{c}}
{0.0035}&{1.4166}\\
{1.4166}&{5908.0319}
\end{array}} \right]
\end{eqnarray*}
This implies that \({\left\| {G\left( s \right)} \right\|_{{L_\infty }}} < 0.9\) is convinced. However, when setting \(\delta {\rm{ = }}0.8\), LMI (27)
cannot be solved because 0.8 is less than the max value shown in Fig. 1. It means that Theorem 4 is correct. Further, by using the mincx solver in MATLAB, it can be got that \({\left\| {G\left( s \right)} \right\|_{{L_\infty }}} = 0.8486\), which is corresponds to the maximum value shown in Fig. 1.
\end{example}

\begin{example}
{One newer area of research which must be mentioned is the application of SFOSs to network theory. {Westerlund et al. first proposed a new linear capacitor model in \cite{S.West:1994TDEI}, and after that, many material scientists have studied and proved the existence of fractional capacitance and other fractional components from different perspectives \cite{elshurafa2013APL,john2017APL,agambayev2017CEC}. The fractional capacitance is based on Curie’s empirical law which states that the current through a capacitor is }
{
\begin{eqnarray*}\label{a1}
i\left( t \right) = \frac{{{u_0}}}{{{h_1}{t^\alpha }}}
\end{eqnarray*}
where $h_1$ and $\alpha$ are constants, $u_0$ is the DC voltage applied at $t=0$, and $0<\alpha <1,(\alpha\in\mathbb{R})$.
For an input voltage $u(t)$, the current is}
{
\begin{eqnarray*}
i(t) = C_0{\mathscr{D}^{\alpha}}u(t),
\end{eqnarray*}
where $C_0$ is the capacitance of the capacitor, which is related to a kind of dielectric. Another constant $\alpha$(order) is related to the loss of the capacitor.}

{Let \(C_0=1, \alpha=0.2\). Applying this special capacitor to the following circuit in Fig. 2(a) and using the equivalent circuit in Fig. 2(b),  a singular fractional order system is obtained as
\begin{eqnarray*}
\left\{ {\begin{array}{*{20}{c}}
{\left[ {\begin{array}{*{20}{c}}
1&0\\
0&0
\end{array}} \right]{\mathscr{D}^{0.2}}x = \left[ {\begin{array}{*{20}{c}}
0&1\\
1&0
\end{array}} \right]x + \left[ {\begin{array}{*{20}{c}}
0\\
1
\end{array}} \right]u},\\
\\
{\ \ y = \left[ {\begin{array}{*{20}{c}}
0&{\beta}R
\end{array}} \right]x,}
\end{array}} \right.
\end{eqnarray*}}
where \(x = {\left[ {\begin{array}{*{20}{c}}
{{u_c}}&{{i_c}}
\end{array}} \right]^{\rm T}}\).
The singularity of the coefficient matrix reflects the fact that unless \({u_c}(0) \hspace{-2pt}= \hspace{-2pt}-u(0)\) and \({i_c}\left( 0 \right) = 0\), there will be an impulse when the circuit is turned on at $t = 0$.

\begin{figure}[!htbp]
\centering
\vspace{-14pt}
\includegraphics[width=1\columnwidth]{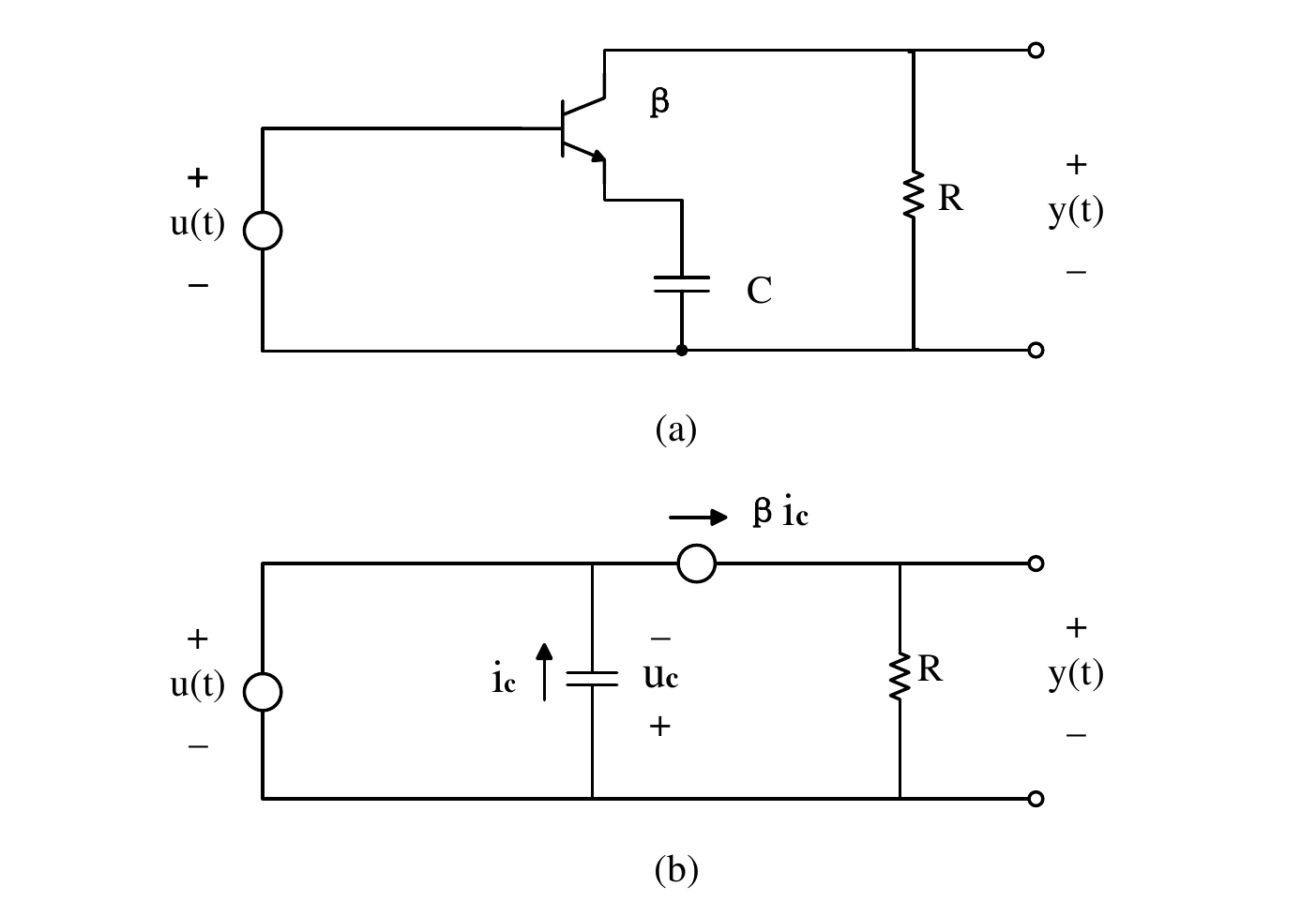}\vspace{-8pt}
\caption{The circuit structure diagram of Example 2.}\label{Fig2}
\end{figure}
}

{{To simplify the calculation, let $R=1/\beta $.}
The maximum singular values of this SFOS are shown in Fig. 3.  At the same time, according to the L-BR lemma at infinite frequency band in Theorem 7, there is \({\left\| {G\left( s \right)} \right\|_{L\infty }}{\rm{ =  + }}\infty \). Therefore, it can be indirectly judged that this system is unstable and the theorem is correct.
\begin{figure}[!htbp]
\centering
\vspace{-10pt}
\includegraphics[width=1\columnwidth]{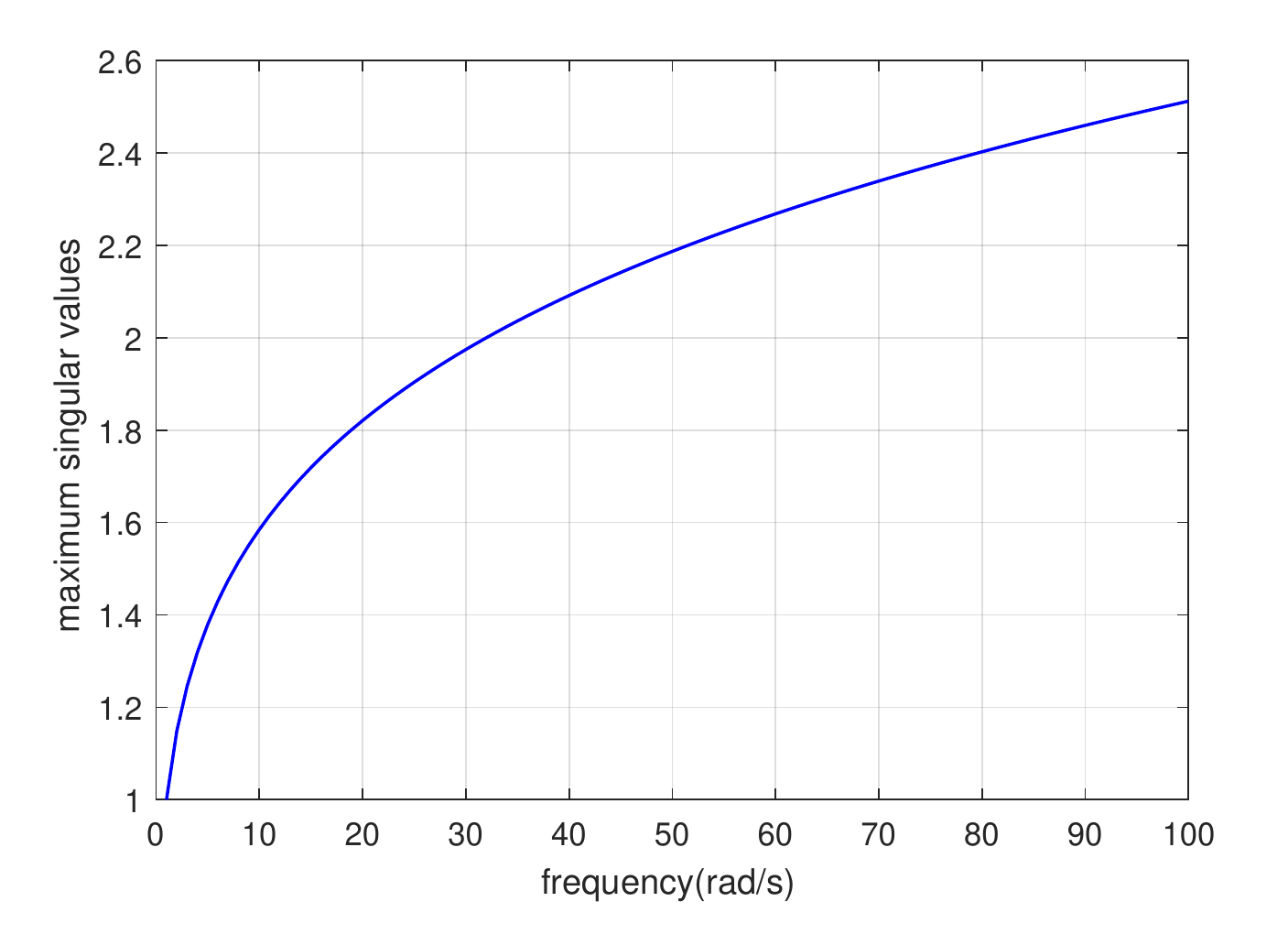}\vspace{-15pt}
\caption{Maximum singular values of the SFOS in Example 2.}\label{Fig3}
\end{figure}
}
\end{example}

\begin{example}
Consider a singular fractional order system with the following state space representation
\begin{eqnarray*}
\left\{ {\begin{array}{*{20}{c}}
{\left[ {\begin{array}{*{20}{c}}
1&0\\
0&0
\end{array}} \right]{\mathscr{D}^{0.5}}x = \left[ {\begin{array}{*{20}{c}}
1&2\\
2&{ - 1}
\end{array}} \right]x + \left[ {\begin{array}{*{20}{c}}
1\\
1
\end{array}} \right]u,}\\
\\
{\ \ \ \ \ \ \ \ \ \ \ y = \left[ {\begin{array}{*{20}{c}}
2&1
\end{array}} \right]x+0.2u.}
\end{array}} \right.
\end{eqnarray*}

The maximum singular values are shown in Fig. 4.
By using the mincx solver in MATLAB, there is \({\left\| {G\left( s \right)} \right\|_{{L_\infty }}} = 2.8971 \).
If the required performance index is given as \({\left\| {G\left( s \right)} \right\|_{{L_\infty }}} < 1 \), then a controller method should be designed to achieve the target.

According to Theorem 8, one has \(K = [\begin{array}{*{20}{c}}
{4.850}&{ - 3.084}
\end{array}]\). After feedback, \({\left\| {G_c\left( s \right)} \right\|_{{L_\infty }}} = 0.7964 \), which meets the requirements of the performance indicators. The maximum singular values of the closed-loop system are shown in Fig. 5.
\begin{figure}[!htbp]
\centering
\includegraphics[width=1\columnwidth]{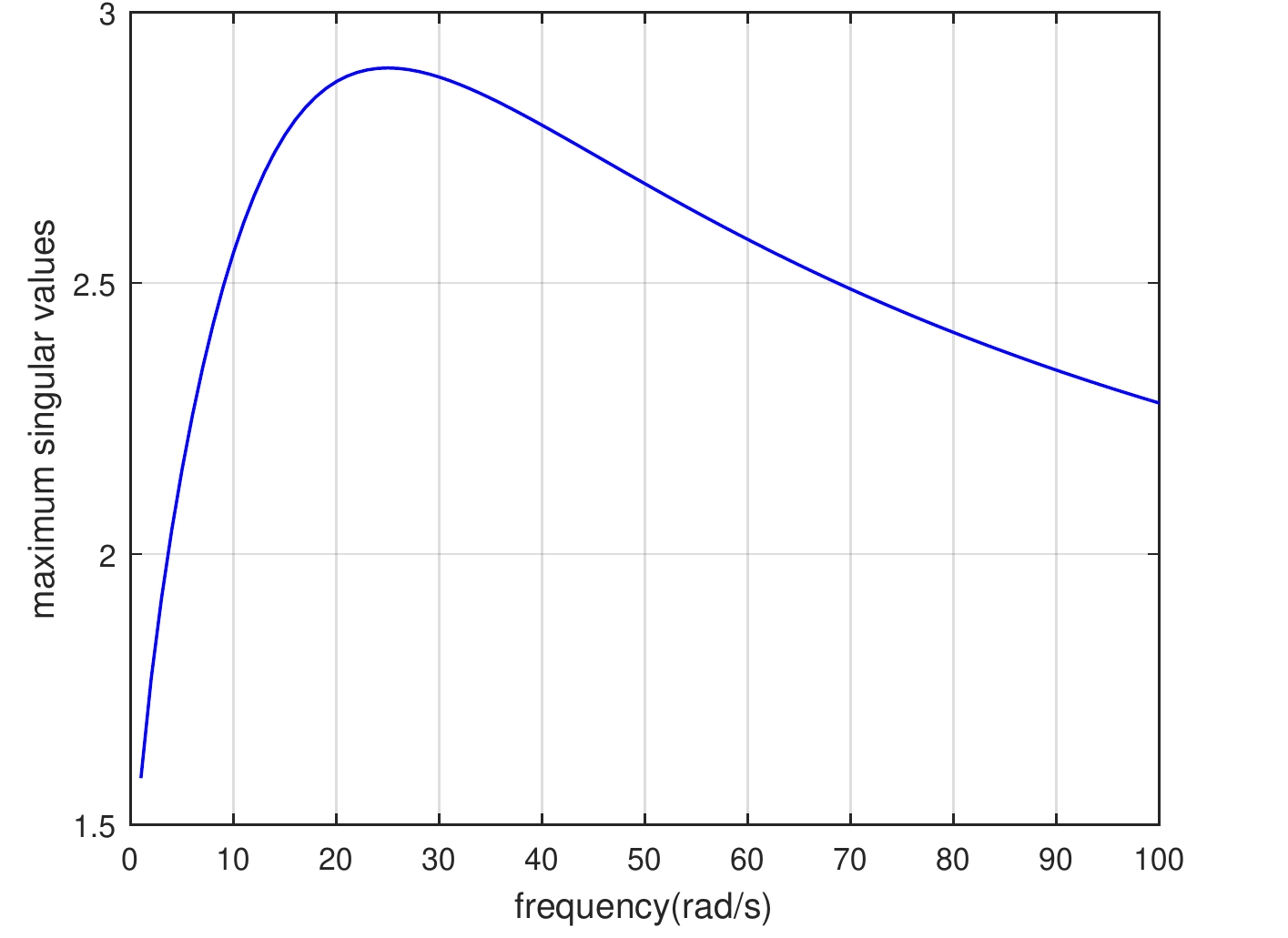}\vspace{-10pt}
\caption{Maximum singular values of the open-loop system in Example 3.}\label{Fig2}
\end{figure}
\begin{figure}[!htbp]
\centering
\includegraphics[width=1\columnwidth]{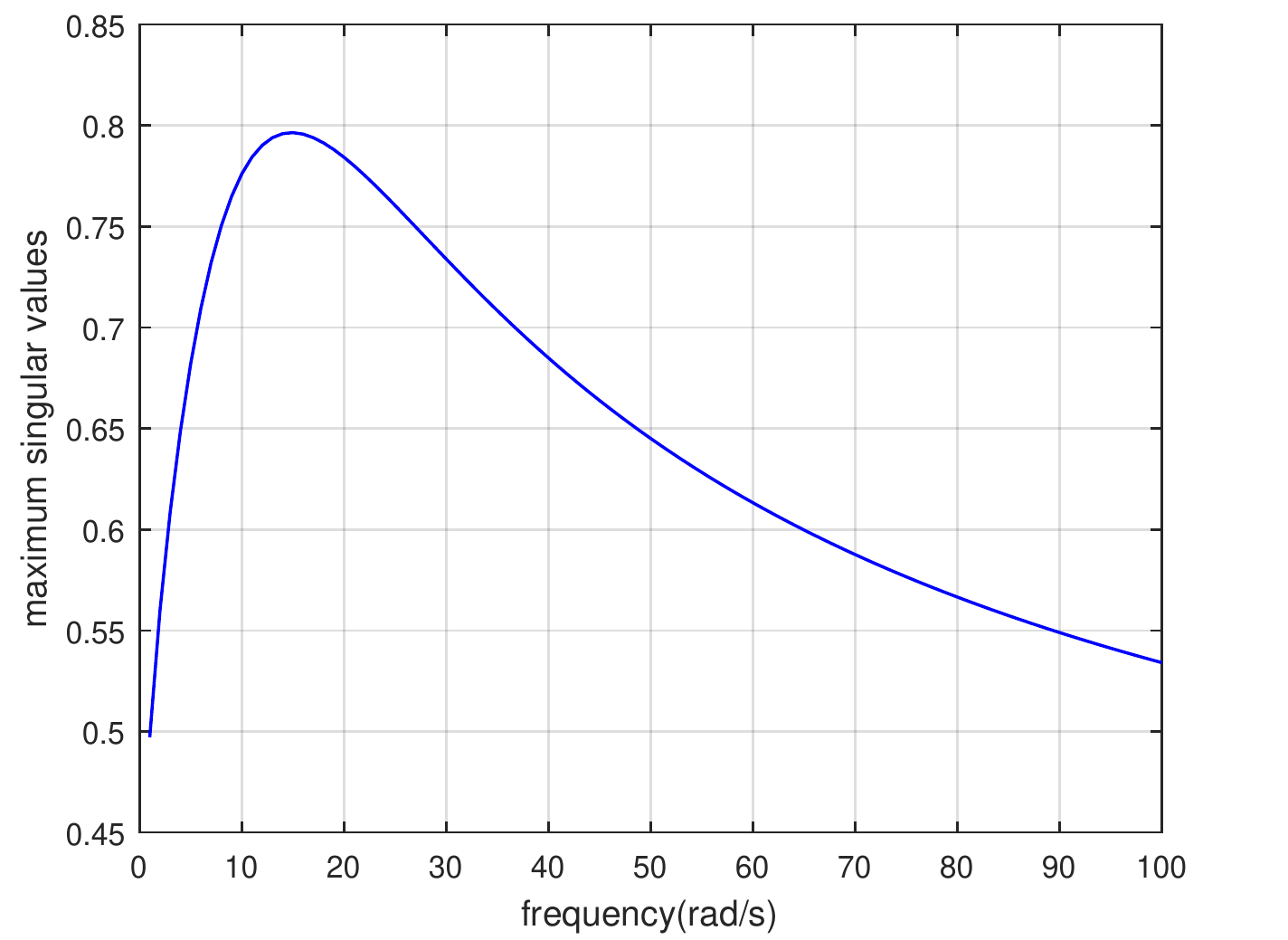}\vspace{-10pt}
\caption{Maximum singular values of the closed-loop system in Example 3.}\label{Fig3}
\end{figure}

\end{example}

\section{Conclusions}\label{Section 7}
In this {paper}, a universal framework of finite frequency band GKYP lemma has been well developed for the SFOS. {This is the first time to fundamentally study the singular fractional-order system in finite frequency ranges without normalization constraints, thus the GKYP lemma on it is brand new.} Based on the proposed GKYP lemma, the $L_\infty$ bounded real lemma of SFOSs in different frequency ranges is obtained, and the controller is designed to effectively satisfy the performance index. {The future research directions include the analysis and synthesis of SFOSs with time-delay, the control and optimization of uncertain SFOSs in different frequency ranges, the passivity of SFOS, the decoupling of large-scale systems' singular representation and the applications of finite frequency SFOSs in power, mechanical and aerospace systems.}

\section*{Acknowledgement}
The work described in this paper was fully supported by the National Natural Science Foundation of China (No. 61601431, No. 61573332), the Anhui Provincial Natural Science Foundation (No. 1708085QF141), the Fundamental Research Funds for the Central Universities (No. WK2100100028) and the General Financial Grant from the China Postdoctoral Science Foundation (No. 2016M602032).

\bibliographystyle{IEEEtran}
\bibliography{IEEEabrv,TACdatabase}

\end{document}